\documentclass[aps,prb,groupedaddress,showkeys,showpacs,amsmath]{revtex4}

\usepackage{amssymb}
\usepackage{amsmath}
\usepackage{amscd}
\usepackage{graphicx}
\usepackage{amsbsy}
\usepackage{color}

\begin{document}
\author{Achille Giacometti}
\affiliation{Dipartimento di Dipartimento di Scienze Molecolari e Nanosistemi, Universit\`a Ca' Foscari Venezia,
Calle Larga S. Marta DD2137, I-30123 Venezia, Italy}
\email{achille@unive.it}
\title{Self-assembly mechanism in colloids: perspectives from Statistical Physics}
\date{\today}

\begin{abstract}
Motivated by recent experimental findings in chemical synthesis of colloidal particles, 
we draw an analogy between self-assembly processes occurring 
in biological systems (e.g. protein folding) and a new exciting possibility in the field of material science.  We consider a self-assembly process 
whose elementary building blocks are decorated patchy colloids of various types, that spontaneously drive the system toward a unique and predetermined targeted 
macroscopic structure.

To this aim, we discuss a simple theoretical model
-- the Kern-Frenkel model -- describing a fluid of colloidal spherical particles with a pre-defined number and distribution
of solvophobic and solvophilic regions on their surface. The solvophobic and solvophilic regions are described via a short-range square-well
and a hard-sphere potentials, respectively.

Integral equation and perturbation theories are presented to discuss structural and thermodynamical properties, with particular emphasis 
on the computation of the fluid-fluid (or gas-liquid) transition in the temperature-density plane.

The model allows the description of both one and two attractive caps, as a function of the fraction of covered 
attractive surface, thus interpolating between a square-well and a hard-sphere fluid, upon changing the coverage. 

By comparison with Monte Carlo simulations, we assess the pros and the cons of both integral equation and perturbation theories in the present
context of patchy colloids, where the computational effort for numerical simulations is rather demanding.
\end{abstract}


\maketitle
\section{Introduction}
\label{sec:intro}
Self-assembly is a process in which components spontaneously form ordered aggregates \cite{Whitesides02}. There exist several examples
of this mechanism, as it is ubiquitous at vary different length and energy scales.

Under appropriate conditions, atoms form crystals that are kept together by strong covalent bonds \cite{Ashcroft76}. This occurs because
the crystal is the ground state of the system with a large energy gap compared with any other disordered aggregate of the same atoms, and
hence the  crystal structure is very robust with respect to any possible defect formation.
The specificity of the interactions and of stoichiometry rules, largely constrains the number of possible final structures that can be obtained,
so that only a relatively small number of crystal structures are found.

At the opposite side stand soft matter systems where interactions are much weaker, of the order of thermal energy \cite{Lyklema91}. 
This includes colloids, polymers, and micelles 
that also can form ordered or disordered macroscopic aggregates under suitable conditions. In this case, however, bonds 
are easily broken and reformed in a solvent, thus allowing the system to rearrange itself in many different ways with configurations separated one another by
small energy barriers. The combined effect of weak interactions and low specificities, hence results in the possibility of obtaining many final aggregates,
typically plagued by many defects and low reproducibility.

Biological systems make an exception to this rule. For instance, a one-dimensional string of amino acids folds reliably and reproducibly into a well
defined unique native state \cite{Lyklema91}, in striking contrast with a similar collapse transition characteristic of a homopolymer,
where the globular structure has many local minima with very small energy differences. The reason for this very different behavior of the polypeptide chain
can be traced back to the specificity of the amino acid sequence combined with the steric effects provided by the side chains \cite{Finkelstein02}.

A similar philosophy has been recently followed in engineering the formation of macroscopic well defined target structure starting
from decorated (patchy) colloids having multiple functionalities and hence specific interactions \cite{Glotzer04,Glotzer07}.
This has become possible in view of the tremendous improvements that have been achieved in the chemical synthesis of colloidal particles
\cite{Walther09,Pawar10}.

While standard colloids display the typical behavior discussed above \cite{Lyklema91}, with many energetically similar aggregated structures,
patchy colloids are also driven by van der Waals-like forces to a spontaneous aggregation in a huge variety of ways, but with only one ground state
that is dictated by the specificity of the anisotropic interactions \cite{Glotzer07}.

A system of patchy colloids, therefore, appears to be one of the most promising route to achieve sufficient control of the design and the interactions
between the building blocks, and hence mimic biological systems such as protein folding \cite{Williamson11}, a very appealing perspective from the
technological view point to create very complex ordered structures at the micro- and nanoscale \cite{Whitesides02}.

The fact that this is indeed a realistic possibility has been recently showed by a set of experiments \cite{Hong08,Chen11}, supported by numerical simulations
\cite{Romano11_a,Romano11_b}.  Here spherical colloidal particles having two hydrophobic poles separated by a charged inner band, were
observed to self-assembly into a two-dimensional well defined ordered structure - not the most natural one associated 
with the symmetry of that particular colloid, in a specific region of the phase diagram.
This will be recalled in Section \ref{sec:self}.

In the remaining of this paper, we will discuss the theoretical approach to this problem in the framework of a specific model of patchy colloids,
the Kern-Frenkel model \cite{Kern03}(see Section \ref{sec:model}), 
that is the same used to numerically reproduce the above experimental findings \cite{Romano11_a,Romano11_b}.
In this paper, we will focus on some theoretical approaches that can capture the essential features of the system behavior, in particular the
part of the phase diagram associated with the fluid-fluid (or gas-liquid) transition.

Two schemes will be discussed, both hinging on classical concepts in statistical physics. 

The first one is an adaptation  to patchy colloids of the usual integral equation theory \cite{Giacometti09a,Giacometti09b,Giacometti10}, 
and is patterned after a similar technique applied to molecular fluids \cite{Lado82,Lado82a,Lado82b,Lado95}. This will be discussed in
Section \ref{subsec:integral}. The second is a thermodynamic perturbation theory, again originally devised for isotropic potentials \cite{Zwanzig54,Barker67},
and more recently applied to the Kern-Frenkel potential \cite{Gogelein08,Gogelein12} (see Section \ref{subsec:perturbation}).
Finally, Section \ref{sec:conclusions} will draw some conclusions and open perspectives.
\section{Self-assembly mechanisms: from protein to patchy colloids}
\label{sec:self}

\subsection{Self-assembly in proteins and in molecules}
\label{subsec:protein}
\begin{figure}[htbp]
\begin{center}
\vskip0.5cm
\includegraphics[width=14cm]{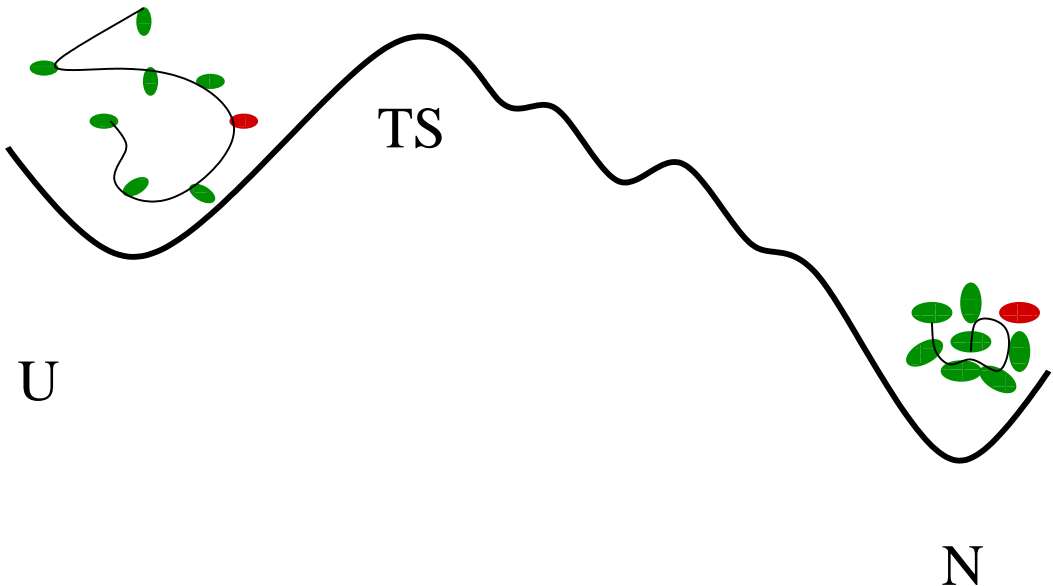}
\caption{A typical folding process. An unfolded polypeptide chain (U) folds into
a well defined native state (N), through a sequence of transitions between
intermediate transitions states (TS) and some local minima. 
\label{fig:fig1}}
\end{center}
\end{figure}

A paradigmatic example of self-assembly process in biology is given by the folding
of a protein \cite{Finkelstein02}. Proteins are biopolymers formed amino acids that are linked together
into a polypeptide chain by covalent bonds. In order to perform the function
that are designed for, each protein must be in its native state, that is in a globular form
with a specific distribution of atoms in space. The protein then spontaneously
folds, following the typical pattern sketched in Fig.\ref{fig:fig1}, from an unfolded (extended)
initial state (U), to a well defined specific
native state (N), where the system is rather compact in space, usually  passing
through one or more transition states (TS), and some local minima.

This self-organization of the protein where a unique target final structure (the native state)
is eventually reached, although the system might pass through a series of
intermediate metastable higher energy states, largely differs from the corresponding
transition in standard polymers. In the polymer case \cite{Doi86} a large number of
possible final globular states, each having comparable, if not identical, energies, are conversely possible.
In both cases, the energy barriers are of the order of the thermal energy and hence a global rearrangement
is possible, and this explains the ``glassy structure'' resulting in the polymer case.
In the protein case, however, the system is driven toward a well defined target structure
by some specificities present in the sequence of the amino acids, and by the steric hindrance given by
side chains that drastically reduce the number of allowed configurations. This allows a global rearrangement
but within a much lower number of local minima, so that the system eventually manages to find its way
toward the native state within biological times (of the order of milliseconds).

Consider now the opposite case of atoms or molecules that assemble into a 
class of well defined crystal structures \cite{Ashcroft76}. In this case, the driving forces are strong
covalent bonds and the final number of possible structures is small because those are the only
ones allowed by the stoichiometry of the interactions. The combination of the strong
interactions (that do not allow for any rearrangement due to thermal energy) and specificity
(given by chemical constraints), forces the system to self-assembly into a unique final structure
compatible with the specificities of all given constraints.
The number of possible crystal structures is however small and enumerated in crystallographic studies.

\subsection{The colloidal domain}
\label{subsec:colloidal}
\begin{figure}[htbp]
\begin{center}
\vskip0.5cm
\includegraphics[width=14cm]{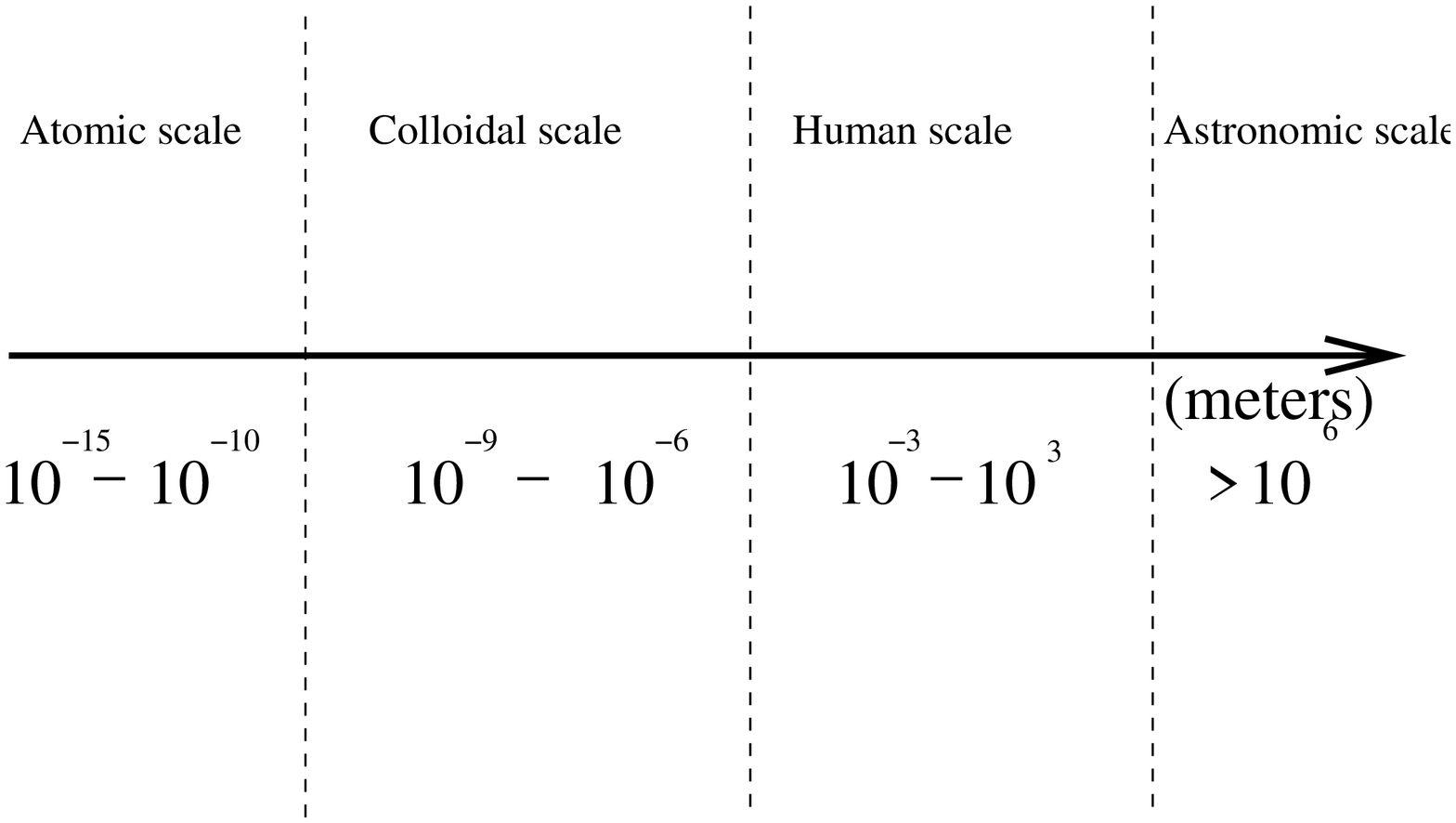}
\caption{The colloidal scale of the order of $10^{-9}$ (nano) and $10^{-6}$ (micro) meters. 
\label{fig:fig2}}
\end{center}
\end{figure}

Colloids are mesoscopic particles with a diameter typically varying between $10^{-8}$ to $10^{-4}$ meters, dispersed
in a microscopic fluid \cite{Lowen94}. The latter is formed by atomistic particles (with dimensions of the order of $10^{-10}$ meters) and
is then often treated as a uniform continuum, so that simple model potentials (and the corresponding phase diagrams)
that are used for atomistic fluids are also valid for colloidal suspensions, but at different length scales,
as sketched in Fig.\ref{fig:fig2}.

One of the main advantage of colloidal suspensions, with respect to their atomistic counterpart, lays in the experimental possibility of
an almost arbitrary tunability in the interactions, in striking contrast with molecular systems where possible interactions are severely
constrained by chemical requirements. An additional attractive feature of colloidal particles, especially from the theoretical point of view, 
is their almost perfect sphericity, so that it is possible to set up a colloidal system behaving as hard-spheres for all practical purposes
\cite{Lyklema91}.

Conversely, the main disadvantage of colloids, not present in atomistic systems, is related to the fact that particles may have slightly
different sizes, charges, or other chemical characteristic, a feature known as polydispersity \cite{Lowen94}.

While this constitutes a problem in the general interpretation of experimental results, several improvements in controlling polydispersity
have been achieved over the years, so that a number of systems now exists where polydispersity can be reduced to values as low as $10\%$ of the
average particle sizes, and the system can be reckoned as monodisperse for almost all practical purposes. 

As other soft matter systems, such as polymers and micelles, colloids have interaction strengths of the order of thermal energy
controlling their organization in clusters. When combined with non specificity in the interactions, this allows for a countless number of possible
complex structures (i.e. target structures), that differ from one another for very tiny free energy differences, that is the
opposite of what one would like to achieve in a mechanism mimicking protein folding.

As we shall see in the next section, this last shortcoming can be now overcome with patchy colloids, where the number of possible target structures
is again huge but, in view of the specificities introduced by the anisotropic potential, they have energy absolute minima well separated by metastable
states having significantly higher energies. Patchy colloids then appear to combine the best of the two worlds, a feature making
these systems very attractive for technological applications, as remarked.  
\subsection{Experimental framework, new chemical synthesis, technological applications}
\label{subsec:experimental}
Chemical synthesis of colloidal particles is far from being new \cite{Lyklema91}, but it has recently experienced
a resurge of interests in view of the remarkable improvement in controlling the composition of the particle surface \cite{Walther09,Pawar10}. 

\begin{figure}[htbp]
\begin{center}
\vskip0.5cm
\includegraphics[width=14cm]{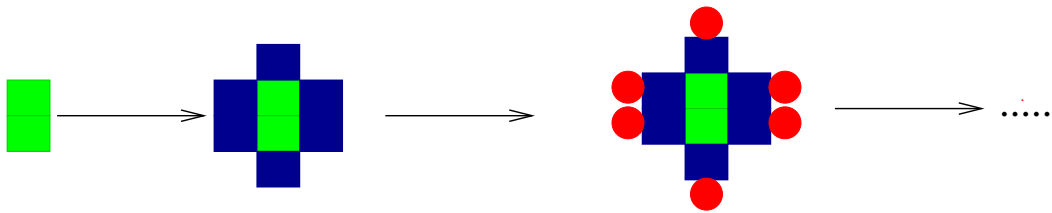}
\caption{A cartoon of the self-assembly process from the elementary units to the final target structure. Different colors and shapes
represent the different building blocks (three in the present case) involved in the process.
\label{fig:fig3}}
\end{center}
\end{figure}

Today, there exists a well defined set of experimental protocols to produce in a controlled way a wide variety of colloidal particles
with different anisotropies in shape, chemical composition, and functionality. 
In particular, it is now possible to directly modify the surface pattern of colloids to produce multiple surface functionality, an experimental
technique known as \textit{templating}. This possibility is extremely appealing for technological applications in such, for instance, crystalline arrays
with complex unit cells could be used to construct materials having photonic band gap in the visible \cite{Hynninen07}.
More generally, this would provide the possibility of designing a bottom-up experiment to achieve a predefined target structure starting
from a suitable set of elementary units, as sketched in Fig.\ref{fig:fig3}.

A glowing example of this possibility is provided by recent experiments by Granick and collaborators \cite{Chen11}, where using
triblock Janus particles, to be defined below,
they succeeded in obtaining a particular planar crystal structure known as Kagome lattice. 

\begin{figure}[htbp]
\begin{center}
\vskip0.5cm
\includegraphics[width=14cm]{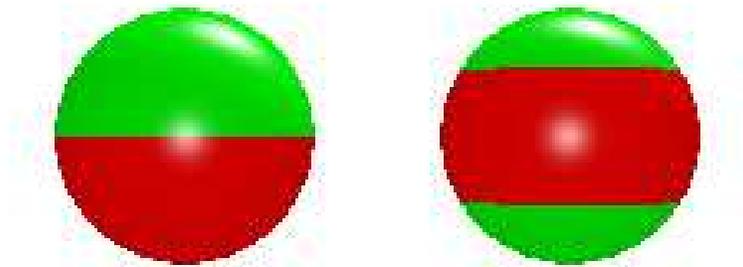}
\caption{The Janus particle (left) and the Janus triblock particle (right). The light and dark regions correspond to the solvophobic and solvophilic regions,
respectively. In both cases, the sum of the solvophobic surface areas equals the solvophilic one.
\label{fig:fig4}}
\end{center}
\end{figure}

The triblock Janus particles (see Fig. \ref{fig:fig4} right) are colloidal particles having the surface partionated
in three parts: two solvophobic caps at the opposite poles of the sphere (the light region of Fig.\ref{fig:fig4}) and one solvophilic internal band (the darker region). 
The name Janus refers to the fact that the sum of the areas associated with the solvophobic and solvophilic regions are almost identical.
The most remarkable feature of this experiment lays in the fact that the triblock Janus particle has not the most natural symmetry
to obtain a two-dimensional Kagome lattice, that requires each particle to have four nearest-neighbors. A more natural symmetry would clearly be a sphere 
with four solvophobic patches. However experimentally this pattern is found to be 
extremely hard to obtain in a reliable and controlled way. The authors of Ref.\onlinecite{Chen11}, supported by successive numerical simulations
\cite{Romano11_a,Romano11_b}, nonetheless found that the
same final target structure can be achieved by using a triblock Janus fluid within a specific region in the pressure-density plane.

In the original Janus particles (Fig.\ref{fig:fig4} left) the surface of the sphere is partionated in only two rather than three parts, and
this has very interesting features on its own right, as we will see in the next section. They can be obtained by using a technique known
as Pickering emulsion, that is a clever way of trapping colloidal particles at the interface of a water-wax emulsion so that they can be 
conveniently functionalized on one side only. The advantage of this technique, compared with traditional templating techniques where
colloidal particles laying on a plane are templated and functionalized on one side, is related to the higher yield in terms of particle production,
due to the lack of the two-dimensional constraint.

Janus particles in pure water tend to repel each other, in view of the strong
solvophilic forces acting on them. However, this repulsion can be screened by adding suitable salt, thus leading to a spontaneous
formation of clusters of various geometries \cite{Hong08}.

Below, we will address a theoretical model what is able to mimic this behavior.
\section{The Kern-Frenkel model of patchy colloids}
\label{sec:model}
The model we are considering is due to Kern-Frenkel \cite{Kern03} and the idea is the following. We consider a fluid
of spheres where the surface of the sphere is divided in two parts having square-well and hard-spheres character, the first
mimicking the solvophobic region, the second the solvophilic, within an implicit solvent description.
 
\begin{figure}[htbp]
\begin{center}
\vskip0.5cm
\includegraphics[width=10cm]{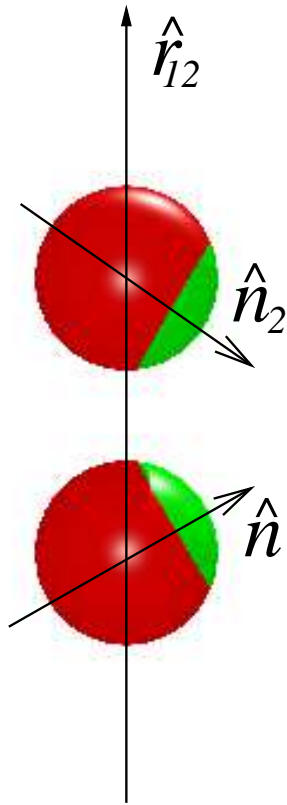} \\
\caption{The one-patch Kern-Frenkel potential, where $\hat{\mathbf{r}}_{12}$ is the direction joining the two centers. 
Directions of the patches are specified by unit vectors $\hat{\mathbf{n}}_1$ and $\hat{\mathbf{n}}_2$ and the surface coverage is $\chi=0.8$.
\label{fig:fig5}}
\end{center}
\end{figure}

The position of each particle in space is identified be a set of vectors $\mathbf{r}_i$, with $i=1,\ldots,N$ $N$ being the total number of particles,
whereas the augular orientation each square-well patch on the surface sphere 
is identified by unit vectors  $\hat{\textbf{n}}_{i}$.  
Finally, the direction connecting centers of
spheres $i$ and $j$ are characterized by unit vector $\hat{\mathbf{r}}_{ij}=(\mathbf{r}_i-\mathbf{r}_j)/ \vert \mathbf{r}_i-\mathbf{r}_j \vert$. 
Figure \ref{fig:fig5} depicts the situation
in the case $i=1$ and $j=2$.

Therefore, two spheres of diameter $\sigma$ attract each other via a square-well potential of width $(\lambda-1) \sigma$ and depth $\epsilon$ if the directions 
of the patches on each sphere
are within a solid angle defined by $\theta_0$ with their relative distance within the range of the attractive well, 
and repel each other as hard spheres otherwise. As the system is still translational invariant, the pair potential depends upon the difference
$\mathbf{r}_{ij}=\mathbf{r}_{i}-\mathbf{r}_{j}$, rather than $\mathbf{r}_{i}$ and $\mathbf{r}_{j}$ separately, and has the form
\begin{eqnarray}
\Phi\left(\mathbf{r}_{ij},\hat{\mathbf{n}}_{i},\hat{\mathbf{n}}_{j}\right) &=& \phi_{\text{HS}} \left(r_{ij}\right) +
\phi_{\text{SW}} \left(r_{ij}\right) \Psi\left(\hat{\mathbf{n}}_{i}, \hat{\mathbf{n}}_{j},\hat{\mathbf{r}}_{ij}\right)
\label{kf:eq1}
\end{eqnarray}

where $r_{ij}=\vert \mathbf{r}_{ij} \vert$. The first term in Eq.(\ref{kf:eq1}) is the hard-sphere contribution
\begin{equation}
\phi_{\text{HS}}\left(r\right)= \left\{ 
\begin{array}{ccc}
\infty,    &  &   0<r< \sigma    \\ 
0,          &  &   \sigma < r \ %
\end{array}%
\right.  \label{kf:eq1b}
\end{equation}
while the second term can be factorized into a isotropic square-well tail
\begin{equation}
\phi_{\text{SW}}\left(r\right)= \left\{ 
\begin{array}{ccc}
- \epsilon, &  &   \sigma<r< \lambda \sigma   \\ 
0,          &  &   \lambda \sigma < r \ %
\end{array}%
\right.  \label{kf:eq2}
\end{equation}
multiplied by an angular dependent factor
\begin{equation}
\Psi\left(\hat{\mathbf{n}}_i,
\hat{\mathbf{n}}_j,\hat{\mathbf{r}}_{ij}\right)= \left\{ 
\begin{array}{ccccc}
1,    & \text{if}  &   \hat{\mathbf{n}}_i \cdot \hat{\mathbf{r}}_{ij} \ge \cos \theta_0 & \text{and} &  
-\hat{\mathbf{n}}_j \cdot \hat{\mathbf{r}}_{ij} \ge \cos \theta_0 \\ 
0,    &  & &\text{otherwise} & \
\end{array}
\right.  \label{kf:eq3}
\end{equation}
The unit vectors $\hat{\mathbf{n}}_{i}(\omega_{i})$, ($i=1,2$) are defined by the 
spherical angles $\omega_i=(\theta_i,\varphi_i)$ in an arbitrarily oriented coordinate frame and $\hat{\mathbf{r}}_{ij}(\Omega)$  
is identified by the spherical 
angle $\Omega$ in the same frame. 
Here $\beta\equiv\left(  k_{\mathrm{B}}T\right)  ^{-1}$, $k_{\mathrm{B}}$ is Boltzmann's constant, and $T$ the absolute temperature.
Reduced units, for temperature $T^*=k_{\mathrm{B}} T/\epsilon$, pressure $P^{*}=\beta P/\rho$, density $\rho^*=\rho \sigma^3$,
and chemical potential $\mu^{*}=\beta \mu$ are routinely used in the description of the thermodynamic. 
The above potential then ensures that, in order to be bonded, two particles have to be within the range of the square-well potential with their attractive surfaces 
properly aligned. If not, they behave as hard spheres. 

The relative ratio between attractive and total  surfaces 
is the  coverage $\chi$ that is related to the semi-angular width $\theta_0$ of the patch. This can be obtained as \cite{Kern03,Giacometti09b}
\begin{eqnarray}
\label{kf:eq4}
\chi &=& \left\langle \Psi\left(\hat{\mathbf{n}}_i,\hat{\mathbf{n}}_j,\hat{\mathbf{r}}_{ij} 
\right) \right \rangle_{\omega_i \omega_j}^{1/2} = \sin^2\left(\frac{\theta_0}{2}\right).
\end{eqnarray}
where we have introduced the angular average
\begin{eqnarray}
\label{kf:eq5}
\left \langle \ldots \right \rangle_{\omega} &=& \frac{1}{4 \pi} \int d \omega \ldots
\end{eqnarray}
\section{The tools of statistical physics}
\label{sec:tools}
As discussed above, it is crucial for self-assembly process to have a full control of the phase behavior of the system to make
sure that a particular condensed phase is obtained for a given set of parameters.

Statistical physics has developed a number of different techniques able to assess thermophysical properties
of simple and complex fluids. Here we will mention two of them that have proven to be useful in the context of self-assembly  patchy colloids.
\subsection{Integral equation theories}
\label{subsec:integral}
In the case of spherically symmetric potentials, the strategy to infer 
thermophysical properties of a fluid of density $\rho$ has a long and venerable
tradition in integral equation theories \cite{Hansen86}. The basic idea is to be able to compute the total correlation function $h(r)$
that is directly linked to all structural and thermodynamical properties of the fluid. In order to do that, however,
the knowledge of the direct correlation function $c(r)$ is also required. The two are related by the exact
Ornstein-Zernike (OZ) equation
\begin{eqnarray}
\label{integral:eq1}
h\left( r\right) =c\left( r\right) +\rho \int d \mathbf{r}^{\prime }~c\left( r^{\prime }\right)
~h\left( |\mathbf{r-r}^{\prime }|\right) 
\end{eqnarray}%
but an additional equation is required in order to close the system and compute the two functions. The second relation
also involves the potential $\phi(r)$, and has the general form
\begin{eqnarray}
\label{integral:eq2}
c\left(r\right) &=& \exp{\left[-\beta \phi\left(r\right) + \gamma\left(r\right)+B \left(r\right) \right]}-1 -\gamma\left(r\right)
 \end{eqnarray}
where $\gamma\left( r\right) =h\left( r\right)-c\left( r\right)$. 

Note that $h(r)=g(r)-1$, $g(r)$ being the radial distribution function that is usually computed in numerical simulations.
The crucial point in this matter, is that this second equation (\ref{integral:eq2}) cannot be computed exactly and an approximation (closure)
is required to compute the bridge function $B(r)$, that is known only as an infinite power series in density whose coefficients cannot 
be readily calculated. All practical closures then approximate $B\left(r\right)$ in some way, and this gives rise to thermodynamic
inconsistencies that are well-known in this formalism \cite{Hansen86}.
The numerical solution of the system of non-linear equations (\ref{integral:eq1}) and (\ref{integral:eq2}) then proceeds through a self-consistent
procedure going back and forth from real to momentum space until self-consistency is achieved, as sketched in Table I. 
\begin{table}
{\large
\begin{equation*}
\begin{CD}
c\left(r\right)
@>\text{Fourier transform} >>
\hat{c}\left(k\right) \\
@AA\text{Closure } A  @.    @VV\text{OZ equation  }V\\
\gamma\left(r\right)
@<\text{Inverse Fourier transform} <<
\hat{\gamma}\left(k\right) \\
\end{CD}
\end{equation*}
}
\label{table:tab1}
\caption{Schematic flow-chart for the solution of the OZ equation
for isotropic potentials.}
\end{table}
Clearly the reason why it proves convenient
to go into momentum space, is related to the fact that the convolution appearing in OZ equation (\ref{integral:eq1}) greatly
simplifies in Fourier form
\begin{eqnarray}
\label{integral:eq3}
\hat{h}\left(k\right) &=& \frac{\hat{c}\left(k\right)}{1-\rho \hat{c}\left(k\right)}
\end{eqnarray} 
$\hat{h}(k)$ and $\hat{c}(k)$ being the Fourier transform of $h(r)$ and $c(r)$ respectively. Use of the auxiliary
function $\gamma(r)$ in place of $h(r)$ is also more convenient in practical calculations, as illustrated in Table I \cite{Labik85}.

The case of angular dependent anisotropic potentials is far more complex from the algorithmic point of view, but the philosophy
behind the methodology is identical. It was devised in the frame of molecular fluids \cite{Gray84}, and more recently 
adapted to the case of patchy colloids \cite{Giacometti09b,Giacometti10}. Here we just sketch the idea, referring to Refs. 
\onlinecite{Giacometti09a,Giacometti09b,Giacometti10} for details. The whole procedure hinges on a remarkable piece of
work carried out by Fred Lado in a series of papers in the framework of molecular fluids \cite{Lado82,Lado82a,Lado82b}.
 
The iterative solution of the angular dependent Ornstein-Zernike (OZ) equation plus an 
approximate closure equation again requires a series of direct and inverse Fourier transforms between real and momentum space involving the
bridge function $B(\mathbf{r}_{ij},\hat{\mathbf{n}}_{i},\hat{\mathbf{n}}_{j})$, the direct correlation function 
$c(\mathbf{r}_{ij},\hat{\mathbf{n}}_{i},\hat{\mathbf{n}}_{j})$, the pair distribution $g(\mathbf{r}_{ij},\hat{\mathbf{n}}_{i},\hat{\mathbf{n}}_{j})$, the correlation functions  $h(\mathbf{r}_{ij},\hat{\mathbf{n}}_{i},\hat{\mathbf{n}}_{j})=g(\mathbf{r}_{ij},\hat{\mathbf{n}}_{i},\hat{\mathbf{n}}_{j})-1$, and the auxiliary function $\gamma(\mathbf{r}_{ij},\hat{\mathbf{n}}_{i},\hat{\mathbf{n}}_{j})=h(\mathbf{r}_{ij},\hat{\mathbf{n}}_{i},\hat{\mathbf{n}}_{j})-c(\mathbf{r}_{ij},\hat{\mathbf{n}}_{i},\hat{\mathbf{n}}_{j})$.
 
Angular dependence introduces additional direct and inverse Clebsch-Gordan transformations \cite{Gray84} 
between the coefficients of angular expansions in ``axial'' 
frames, with $\hat{\mathbf{z}}= \hat{\mathbf{r}}_{ij}$ in direct space or $\hat{\mathbf{z}} = \hat{\mathbf{k}}$ in momentum space, and more general ``space'' frames with arbitrarily-oriented axes. In the case of the Kern-Frenkel potential of Section \ref{sec:model} with only one or two patches, 
the cylindrical symmetry of the angular 
dependence entitles the use of the simpler version of the procedure for linear molecules \cite{Lado82a} rather than the full anisotropic version \cite{Lado95}.
The results is illustrated in Table II and is the extension of the isotropic case given in Table I. 
\begin{table} 
{\large
\begin{equation*} 
\begin{CD}
c\left(r;l_{1}l_{2}l\right)
@>\text{Hankel transform }>>
\tilde{c}\left(k;l_{1}l_{2}l\right)
@>\text{Inverse CG transform }>>
\tilde{c}_{l_{1}l_{2}m}\left(k\right) \\
@AA\text{CG transform } A  @.    @VV\text{OZ equation  }V\\
c_{l_{1}l_{2}m}\left(r\right)
@.      @.
\tilde{\gamma}_{l_{1}l_{2}m}\left(k\right) \\     
@AA\text{Expansion inverse} A  @.    @VV\text{CG transform }V\\
c\left(r,\omega_1,\omega_2\right)
@.      @.
\tilde{\gamma}\left(k;l_{1}l_{2}l\right) \\
@AA\text{Closure } A  @.    @VV\text{Inverse Hankel transform }V\\
\gamma\left(r,\omega_1,\omega_2\right)   
@.      @.
\gamma\left(r;l_{1} l_{2} l\right) \\
@AA\text{Expansion } A   @.   @VV\text{Inverse CG transform  }V\\
\left[\gamma_{l_{1} l_{2} m}\left(r\right) \right]_{\text{old}}   
@<<<    \text{Iterate}    @<<<  
\left[\gamma_{l_{1} l_{2} m}\left(r\right) \right]_{\text{new}}
\end{CD}
\end{equation*}
}
\label{table:tab2}
\caption{Schematic flow-chart for the solution of the OZ equation
for the Kern-Frenkel angle-dependent potential.See Section \ref{subsec:integral} for a description of the scheme}
\end{table}
For $i=1$ and $j=2$, the procedure in real space starts an iteration in the axial frame with the current values of the axial coefficients $\gamma_{l_{1} l_{2} m}(r_{12})$ 
and uses an expansion in spherical harmonics to construct $\gamma(r_{12},\omega_1,\omega_2)$. 
The bridge function $B(r_{12},\omega_1,\omega_2)$ is similarly constructed in a given closure approximation, and the latter is used to obtain 
$c(r_{12},\omega_1,\omega_2)$, its axial coefficients $c_{l_{1} l_{2} m}(r_{12})$ and, through a Clebsch-Gordan (CG) transformation, 
the corresponding space coefficients $c(r_{12};l_{1} l_{2} l )$. 

The corresponding part in Fourier space is achieved by Hankel transform, and a backward Clebsch-Gordan transformation 
to return to an axial frame in $\mathbf{k}$ space, yielding $\tilde{c}_{l_{1}l_{2} m}(k)$. A parallel sequence of operations
in Fourier space starts with the OZ equation to get $\tilde{\gamma}_{l_{1} l_{2} m}(k)$, followed by a forward Clebsch-Gordan 
transformation, and an inverse Hankel transform, to provide $\gamma(r_{12};l_{1} l_{2} l)$. 
A final backward Clebsch-Gordan transformation, brings a new estimate of the original coefficients $\gamma_{l_{1} l_{2} m}(r_{12})$. 
This cycle is repeated until self-consistency between input and output $\gamma_{l_{1} l_{2} m}(r_{12})$ is achieved as before. 

In the present framework a particular closure denoted as reference hypernetted-chain (RHNC) has been used for both the square-well 
and the Kern-Frenkel potential, with the bridge function approximated by its hard-sphere functional form. 
It has few shortcomings and several advantages, compared to other possible closures, that have
been analyzed in detail in past work \cite{Lado82,Lado82a,Lado82b}. Its main advantage here is related to
its ability of computing the chemical potential and pressure without invoking additional approximations besides 
those included in the closure \cite{Giacometti09a,Giacometti09b,Giacometti10}. For a fixed temperature $T^{*}$, one
can then compute the pressure of the gas (colloidal poor) phase $P_{g}^{*}$ and of the liquid (colloidal rich) $P_{l}^{*}$ phases,
and the corresponding chemical potentials $\mu_{g}^{*}$ and $\mu_{l}^{*}$. The fluid-fluid (gas-liquid) coexistence line then follows
from a numerical solution of a system of non-linear equations 
\begin{eqnarray}
\label{integral:eq4a}
P_{g}^{*} \left(T^{*},\rho^{*}_{g} \right)&=& P_{l}^{*} \left(T^{*},\rho^{*}_{l} \right) \\
\label{integral:eq4b}
\mu_{g}^{*} \left(T^{*},\rho^{*}_{g} \right)&=& \mu_{l}^{*} \left(T^{*},\rho^{*}_{l} \right)
\end{eqnarray} 
whose solutions are the gas $\rho_{g}^{*}$ and liquid $\rho_{l}^{*}$ reduced densities associated with the coexistence lines. 

\begin{figure}[htbp]
\begin{center}
\vskip0.5cm
\includegraphics[width=14cm]{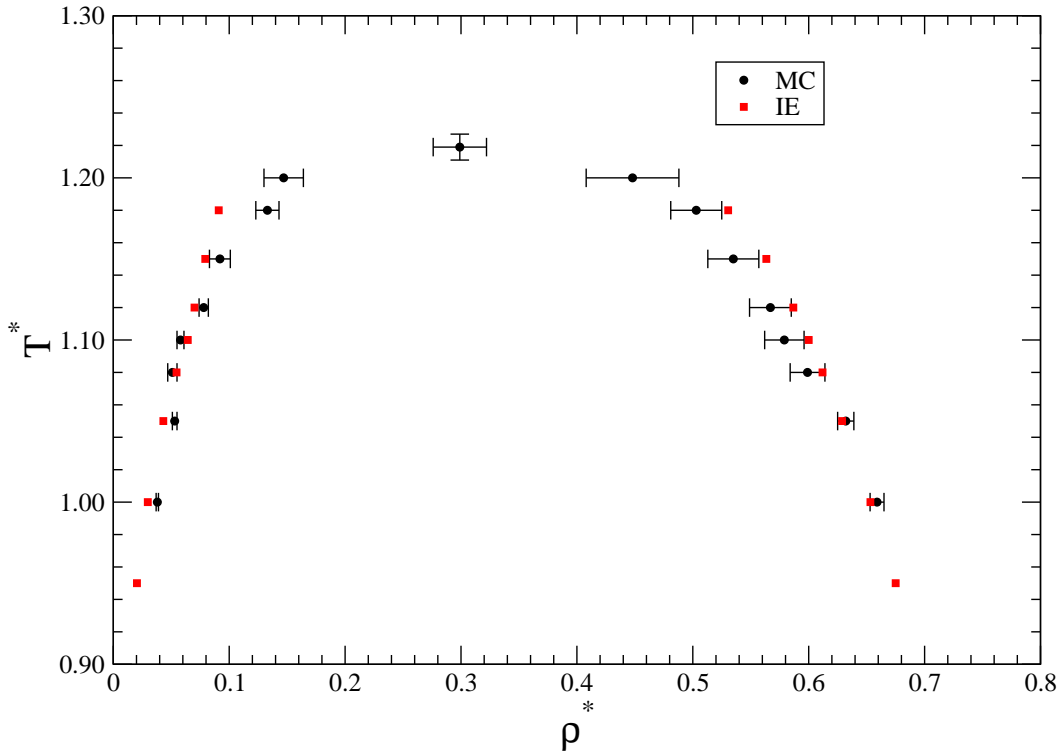}
\caption{The computed phase diagram of a square-well fluid with $\lambda=1.5$. Results from integral equation theory (IE) 
are contrasted with Monte Carlo simulations (MC) of Ref. \onlinecite{Vega92}.
\label{fig:fig6}}
\end{center}
\end{figure}

In case of square-well with a width amplitude of $\lambda=1.5$, the resulting phase diagram in the temperature-density plane is reported in Fig. \ref{fig:fig6} . 

The results from reference
hypernetted-chain integral equation theory (IE) are contrasted with those stemming from Gibbs ensemble Monte Carlo simulations
by Vega \textit{et al} \cite{Vega92} showing good agreement in both branches, far from the critical region. The difficulties 
in assessing the critical region, is a well-known feature of all integral equation theories, and it is related to the aforementioned
thermodynamical inconsistencies related to the closure. Apart from this, the performance of integral equation theory is
rather satisfactory, with the agreement with numerical simulations not only qualitative but also quantitative. This suggests that
the the theory is able to capture the essential features of the phase behavior.
It is also worth recalling that the fluid-fluid transition is known to become metastable against fluid-solid transition below $\lambda=1.25$,
that is for sufficiently short-range attractions \cite{Liu05}. 

\begin{figure}[htbp]
\begin{center}
\vskip0.5cm
\includegraphics[width=14cm]{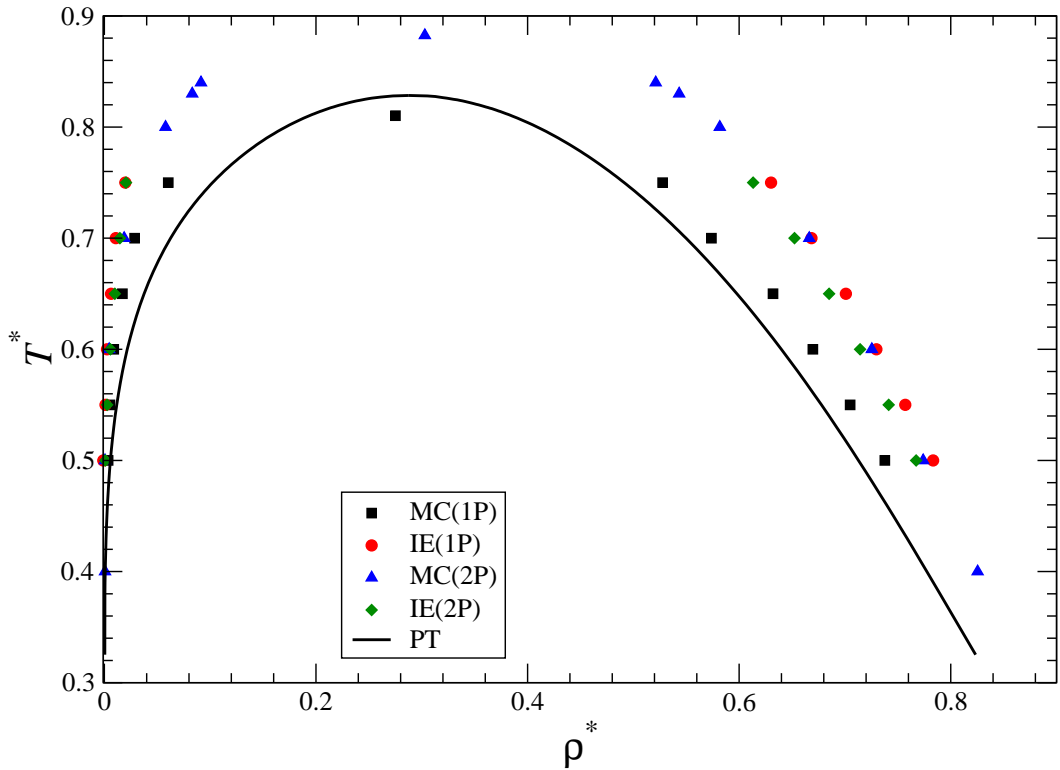}
\caption{The computed phase diagram of a Kern-Frenkel fluid with $\lambda=1.5$ and surface coverage $\chi=0.8$. Results from integral equation theory (IE) are
contrasted with Monte Carlo simulations (MC) obtained
in Ref. \onlinecite{Giacometti09b} and \onlinecite{Giacometti10} both in the case of a single (1P) and double (2P) patches.
The solid line depicts the results from  Barker-Henderson perturbation theory (PT) as given in Ref.\onlinecite{Gogelein12}.
\label{fig:fig7}}
\end{center}
\end{figure}

The case of patchy colloids is tackled using the same methodology \cite{Giacometti09b,Giacometti10}.
Again the  reference hypernetted-chain integral equation theory (IE) is contrasted with MC numerical
simulations by Sciortino \textit{et al} \cite{Sciortino09,Sciortino10} in Fig. \ref{fig:fig7}. 
As representative value, a coverage of $\chi=0.8$ with $\lambda=1.5$ was selected, so this
value is intermediate between the isotropic square-well discussed above, and the Janus limit corresponding to half coverage ($\chi=0.5$) that
is known to display an anomalous phase diagram \cite{Sciortino09,Sciortino10}. Unlike the isotropic potential, however, patchy systems allow to
have the same attractive part distributed in different ways on the surface, and this is included in the two cases also reported in Fig.\ref{fig:fig7}. 

The first case is when
all attractive parts are concentrated into a single patch (1P), while the second case corresponds to an attraction distributed into
two circular patches at the opposite poles of the sphere (2P). Note that, at half coverage $\chi=0.5$, the two cases correspond to the Janus and the
triblock Janus particle, respectively. The phase diagram of the Janus case turns out to be particularly interesting, as it is ruled
by a competing mechanism with the tendency to form a condensed phase on the one hand, and a clusterization process tending to
form isolated and non-interacting micelles (and vesicles) inhibiting the condensation process, on the other hand. The result is
an anomalous phase diagram with reentrant vaporization \cite{Sciortino09}.

Two features of Fig. \ref{fig:fig7} are noteworthy. First, the numerical simulations clearly indicate that the coexistence line of the
two patches case always lay above the single patch counterpart at equal coverage. This means that, on cooling from the high temperature single
phase region, the two patches system tends to phase separate before the corresponding one patch case, as it was to be expected in view
of the higher valence in the 2P case. Second, the consistency between integral equation theory and numerical simulations
appears to be significantly better in the 2P as compared with the 1P counterpart. Again, this is not surprising in view of the
fact that the one-patch potential is much more asymmetric and hence harder to be described by an isotropic approximation of the bridge function,
such as that used in the present approach.

For sufficiently low coverages, one might expect the fluid-fluid transition to become metastable against other transitions, thus providing
the angular analog of the metastability occurring for short-range attractive square-well interactions discussed previously. This is indeed the case.
For the one-patch potential, extensive simulations \cite{Sciortino09,Sciortino10} indicate that a micellization process prevents the
condensation process altogether. In the two-patches case, on the other hand, even at low coverages the double valence on each particle allows
the formation of a network, so that the condensation process can take place. However, a crystalization was shown to occur for coverages below $\chi \approx 0.3$, 
so that the system forms a crystal before being able to condensate \cite{Giacometti10}.
\subsection{Barker-Henderson perturbation theory}
\label{subsec:perturbation}

The idea behind perturbation theory is that this second term in Eq.(\ref{kf:eq1}) can be treated as a perturbation of the first, that is of the hard-sphere term.
In the present context, this means that we are starting from a fluid of hard-spheres and perturbatively decorating the hard-sphere surfaces
with a attractive patches. We now discuss how a classic perturbation theory due to Barker and Henderson \cite{Barker67,Henderson71,Barker76}
can be adapted to the Kern-Frenkel model  to infer thermodynamical properties  \cite{Gogelein08,Gogelein12}.

In the case of isotropic potential, the original method is due to Zwanzig \cite{Zwanzig54}, but the most useful version was given
by Barker and Henderson few years later \cite{Barker67}. The method was recently extended to the Kern-Frenkel potential 
\cite{Gogelein08,Gogelein12}, and it is based on a second-order cumulant expansion of the excess free energy

In order to illustrate the method, let us derive the first order term. Let us work in the canonical ensemble, although 
it proves much more convenient  to work in  the 
the grand-canonical ensemble, as it correctly takes care of the finite size contributions \cite{Henderson71}. 

Assume the total potential $U$ to have the general following form
\begin{eqnarray}
\label{perturbation:eq1}
U_{\gamma}\left(1,\ldots,N\right)&=& U_{0}\left(1,\ldots,N\right)+\gamma U_{I}\left(1,\ldots,N\right)\\ \nonumber 
&=&\sum_{i<j} \Phi_{\gamma}\left(ij\right) = \sum_{i<j} \Phi_{0}\left(ij\right)+\gamma \sum_{i<j} \Phi_{\text{I}}\left(ij\right)\,\mbox{,}
\end{eqnarray}
where $U_{0}(1,\ldots , N)=\sum_{i,j} \Phi_{0}(ij)$ is the unperturbed part and $U_{I}(1,\ldots ,N)=\sum_{i,j} \Phi_{\text{I}}(ij)$
is the perturbation part. Here $0\le \gamma \le 1$ is used as perturbative parameter. Note that when each coordinate $i$ includes both
the coordinate $\mathbf{r}_i$ and patch orientation $\hat{\mathbf{n}}_i$, so that  $i\equiv (\mathbf{r}_i,\hat{\mathbf{n}}_i)$,
the expression is also valid for the Kern-Frenkel model \cite{Gogelein08,Gogelein12}. For simple fluids, instead $i \equiv \mathbf{r}_i$ only.

The potential represented in Eq.(\ref{kf:eq1}) can be clearly cast in this form upon combining the two parts of the second term appearing in Eq.(\ref{kf:eq1})
\begin{eqnarray}
\label{perturbation:eq2}
\Phi_{I}\left(\mathbf{r}_{ij},\hat{\mathbf{n}}_{i},\hat{\mathbf{n}}_{j}\right)&=& \phi_{\text{SW}} \left(r_{ij}\right) 
\Psi\left(\hat{\mathbf{n}}_{i},\hat{\mathbf{n}}_{j},\hat{\mathbf{r}}_{ij}\right)
\end{eqnarray}

Introducing the following short-hand notation 
\begin{eqnarray}
\label{perturbation:eq3}
\int_{1,\ldots,N} \left(\cdots \right) &\equiv & \int \left[\prod_{i=1}^N d \mathbf{r}_i \left \langle \left( \cdots \right) 
\right \rangle_{\omega_{i}} \right]
\end{eqnarray}
for the integration over all particle coordinates, the partition function 
\begin{eqnarray}
\label{perturbation:eq4}
Q_{\gamma} &=& \frac{1}{N!\Lambda_T^{3N}} \int_{1,\ldots,N} e^{-\beta U_{\gamma}}= \frac{1}{N!\Lambda_T^{3N}} Z_{\gamma}=
e^{-\beta F_{\gamma}}
\end{eqnarray}
(here $\Lambda_T$ is the de Broglie thermal wavelength, and $Z_{\gamma}$ is the configurational partition function), 
can then be used to obtain an expansion of the Helmholtz free energy \cite{Henderson71}.
\begin{eqnarray}
\label{perturbation:eq5}
F_{\gamma} &=& F_0+ \gamma \left(\frac{\partial F_{\gamma}}{\partial \gamma} \right)_{\gamma=0}+
\frac{1}{2!} \gamma^2 \left(\frac{\partial^2 F_{\gamma}}{\partial \gamma^2} \right)_{\gamma=0}+ \cdots
\end{eqnarray}
that is valid for arbitrary $\gamma$.

Taking the derivative of $F_{\gamma}$ one has, using Eq.(\ref{perturbation:eq2})
\begin{eqnarray}
\label{perturbation:eq6}
\frac{\partial}{\partial \gamma} \left[\beta F_{\gamma} \right] &=& \frac{1}{2} \int_{1,2} \frac{\partial}{\partial \gamma}
\left[-\beta \Phi_{\gamma}  \left(12\right) \right] \rho_{\gamma} \left(12\right)\,\mbox{,}
\end{eqnarray}
where
\begin{eqnarray}
\label{perturbation:eq7}
\rho_{\gamma} \left(1 2 \right)&=& \frac{N!}{\left(N-2\right)!}  \frac{1}{Z_{\gamma}} \int_{3,\ldots,N} e^{-\beta U_{\gamma}}\,\mbox{.}
\end{eqnarray}
When $\gamma=1$, this yields the free energy correct to first order. 

An explicit expression of the second order term is far more involved and can be found in Ref.\onlinecite{Gogelein12}.
In the particular case of the Kern-Frenkel potential, one then obtains \cite{Gogelein08,Gogelein12} the
excess free energy per particle with respect to the corresponding hard-sphere free energy $F_{\text{HS}}/N$ to have the reduced form 
\begin{eqnarray}
\label{perturbation:eq8}
\frac{\beta \left(F-F_{\text{HS}}\right)}{N} &=& \frac{\beta F_1}{N}+\frac{\beta F_2}{N}+ \ldots
\end{eqnarray}
with
\begin{eqnarray}
\label{perturbation:eq9}
\frac{\beta F_1}{N} &=& \frac{12 \eta}{\sigma^3} \int_{\sigma}^{\lambda \sigma}  dr r^2  \phi_{\text{SW}} \left(r\right) g_0\left(r\right) \left \langle
\beta \Psi\left(12\right) \right \rangle_{\omega_1,\omega_2}\,\mbox{.}
\end{eqnarray}
and
\begin{eqnarray}
\label{perturbation:eq10}
\frac{\beta F_2}{N} &=& -\frac{6 \eta}{\sigma^3}  \left(\frac{\partial \eta}{\partial P_0^{*}} \right)_T \int_{\sigma}^{\lambda \sigma} dr r^2 
g_0\left(r\right) \phi_{\text{SW}}^2 \left(r\right)
\left \langle \left[\beta \Psi\left(12\right) \right]^2 \right \rangle_{\omega_1,\omega_2}\,\mbox{,}
\end{eqnarray}
Here $P_0^*=\beta P_0/\rho$ is the reduced pressure of the HS reference system, 
and $g_0(r)$ the corresponding radial distribution function. We have also introduced the packing fraction $\eta=\pi \rho \sigma^3/6$.
  
Once that the free energy is known, pressure and chemical potential can be obtained
from the following standard thermodynamical relations \cite{Hansen86}
\begin{eqnarray}
\label{perturbation:eq11a}
\frac{\beta P}{\rho} &=& \eta \frac{\partial}{\partial \eta} \left(\frac{\beta F}{N} \right) \\
\label{perturbation:eq11b}
\beta \mu &=& \frac{\partial}{\partial \eta} \left(\eta \frac{\beta F}{N} \right)\,\mbox{.}
\end{eqnarray}
so that the same procedure given in Eqs.(\ref{integral:eq4a}) and (\ref{integral:eq4b})
can then be applied to infer the phase diagram in the temperature-density plane.

In spite of the fact that this is a high temperature expansion - terms $\beta F_1/N$ and  $\beta F_2/N$
are proportional to $1/T^{*}$ and $(1/T^{*})^2$ respectively, this scheme is found to provide a remarkably good description of the fluid-fluid transition.
This is displayed again in Fig.\ref{fig:fig7}, where the fluid-fluid coexistence curves computed using Barker-Henderson perturbation theory (PT)
again in the case $\lambda=1.5$ and $\chi=0.8$ (that is the same case as before) are reported and compared with previous results from integral equation theory
(IE) and Montecarlo simulations (MC). 

Given the second-order truncation in the high temperature expansion, the performance of the perturbation theory is rather remarkable. It is worth stressing, 
however, that this was also shown to be the case even in the fully isotropic square-well potential \cite{Henderson80}. An additional strong point of this
approach, presented in Ref.\onlinecite{Gogelein12}, is that even the fluid-solid coexistence phase is found to be well reproduced by the Barker-Henderson method.
In this latter respect, an interesting additional improvement could stem from a methodology akin to that presented in Ref.\onlinecite{Likos94},
\section{Conclusions and open perspectives}
\label{sec:conclusions}
In the present paper, we have discussed the phase behavior of a particular fluid of patchy colloids and its potential applications to self-assembly
processes at micro- and nanoscale, of biological inspiration.

Interactions in patchy colloids were described by the Kern-Frenkel model where circular attractive square-well patches decorate the surfaces of hard-spheres.
This model then smoothly interpolates between two well-known and studied isotropic fluids -- a hard-sphere fluid (coverage $\chi=0$) and a square-well
fluid (coverage $\chi=1.0$). It is sufficiently rich to display an interesting behavior, and sufficiently simple to be amenable to a detailed theoretical
treatment.

We have discussed the crucial importance of having a complete control of the phase diagram of the system, including the exact location of the fluid-fluid
(gas-liquid) and fluid-solid coexistence lines. Compare to standard isotropic colloids, the phase diagram of patchy colloids may be far more complex,
and this was hinted in the simple case of the Kern-Frenkel potential, where the attractive part covers only a fraction of the
total surface, that can also be distributed among different parts of the surface. This was specifically addressed when
the entire square-well part was condensed into a single region (one-patch) or partioned in two parts at the opposite poles of a sphere (two-patches).

In order to tackle the calculation of the fluid-fluid transition, we have briefly outlined two strategies 
based on two well-known statistical physics tools, namely integral equation theories and perturbation theory.

In the case of integral equations, the reference hypernetted chain (RHNC) has been used to compute the phase diagram in the temperature-density plane,
with the fluid-fluid coexistence lines following very closely those from numerical Monte Carlo simulations both in the single and double patches case
for a given width of the square-well and a given surface coverage. Integral equation theory has two main strengths. First, they can
be formulated in both the single and double patches case with only minor modifications; second it is quantitatively predictive at a very limited computational
effort as compared to the corresponding Monte Carlo simulations that instead require a major computational effort.
The main drawback of the method is related to its internal thermodynamical inconsistencies preventing the access to the critical region, that 
has then to be extrapolated in some way.

Thermodynamic perturbation theory, on the other hand, does not suffer from this drawback, as the full critical region, including the critical point,
can be evaluated. It has however a different disadvantage that we have tried to underline.
Our formulation builds upon a classic second-order theory  due to Barker and Henderson for simple fluids, and we have discussed  
how this can be adapted to the Kern-Frenkel potential in the case of a single patch.
Again by means of a direct comparison with Monte Carlo simulations on the same model, we have assessed the surprising potentialities of this technique
that go beyond expectation, and its good performance in its predictions makes it a method competitive with integral equation theory.
In this case, the main drawback hinges on the fact that, in its present formulation, the exact location of the patches position is irrelevant,
and therefore cannot be used to discriminate between the single or double patches scenario.

While in principle other similar theoretical methods such as the Weeks-Chandler-Anderson (WCA) perturbation theory
\cite{Weeks71,Andersen76,Chandler83} or the reference interaction site model (RISM) \cite{Chandler72,Hansen86} could be possible, in practise
those described in the present work and their possible improvements \cite{Gray84}, are expected to be the most efficient ones, in view of the particular all-or-none angular
dependence of the attractive part. 

The topic of patchy colloids treated in the present paper is a very active research topic, and many different approaches have been envisaged to tackle the analysis of the
phase behavior in details. Although an exhaustive list of them is clearly beyond the aim of the present work - it can be found in Ref. \onlinecite{Bianchi11},
we would still like to mention some very recent results as they are relevant for the point of the present
discussion. Fantoni \textit{et al} \cite{Fantoni11} used a cluster theory to rationalize micelles formation in the Janus fluid. Reinhardt \textit{et al}
\cite{Reinhardt11} mapped the effect of the patchy colloid anisotropy into a binary mixture of isotropic colloids to explain the Janus anomalous
phase behavior observed in Ref.\onlinecite{Sciortino09}.
A different model with similar aims has also been studied by Tavares \textit{et al} \cite{Tavares10}. In this model, earlier proposed by Bianchi et al
\cite{Bianchi06}, attractive patches are replaced by sticky spots distributed over hard-spheres surfaces. Here, sticky spots act as a specific binding
interactions so that the ground state has all possible favorable contacts saturated, with hence an energy significantly lower with respect
to the other metastable states. However, the point-like nature of the attractive spots makes a local rearrangement rather difficult, and the system
has then the tendency to be trapped into a metastable state, in spite of the large energy gap, a drawback not present in the Kern-Frenkel potential,
where the number, the extension, and the location of the patches can be freely and independently tuned to create the most favorable conditions.
\begin{acknowledgments}
The results presented in this paper have been obtained during enjoyable collaborations with Francesco Sciortino, Fred Lado, Giorgio Pastore, Christoph G\"ogelein,
Julio Largo, and Flavio Romano. 
\end{acknowledgments}
\bibliographystyle{apsrev}

\begin{thebibliography}{99}

\bibitem{Whitesides02} G. M. Whitesides and M. Boncheva, Proc. Natl. Acad. Sci. \textbf{99}, 4769 (2002); G. M. Whitesides and 
B. Grzybowski, Science \textbf{295}, 2418 (2002)

\bibitem{Ashcroft76} N.W. Ashcroft and N. D. Mermin, \textit{Solid State Physics}, (Thomson Learning 1976)

\bibitem{Lyklema91} J. Lyklema, \textit{Fundamentals of Interface and Colloid Science, Vol. I: Fundamentals} (Academic, London, 1991).

\bibitem{Finkelstein02} A. V. Finkelstein and O. B. Ptitsyn \textit{Protein Physics} (Academic Press 2002)

\bibitem{Glotzer04} S. C. Glotzer, Science \textbf{306}, 419 (2004).

\bibitem{Glotzer07} S. C. Glotzer and M. J. Solomon, Nature Mater. \textbf{6}, 557 (2007).

\bibitem{Walther09} A. Walther and A. H. E. M\"uller, Soft Matter \textbf{4}, 663 (2008).

\bibitem{Pawar10} A. B. Pawar and I. Kretzchmar, Macromol. Rapid Commun \textbf{31}, 150 (2010)

\bibitem{Williamson11} A. J. Williamson, A. W. Wilber, J. P. K. Doyle, and A. A. Louis, Soft Matter \textbf{7}, 3423 (2011)

\bibitem{Hong08} L. Hong, A. Cacciuto, E. Luijten , and S. Granick,  Langmuir \textbf{24}, 621 (2008).

\bibitem{Chen11} Q. Chen, S. C. Bae and S. Granick, Nature \textbf{469}, 382 (2011)

\bibitem{Romano11_a} F. Romano and F. Sciortino, Nature Materials \textbf{10}, 171 (2011)

\bibitem{Romano11_b} F. Romano, and F. Sciortino, Soft Matter \textbf{7}, 5799 (2011)

\bibitem{Kern03} N. Kern and D. Frenkel, J. Chem. Phys. \textbf{118}, 9882 (2003).

\bibitem{Giacometti09a} A. Giacometti, G. Pastore, and F. Lado, Mol. Phys. \textbf{107}, 555  (2009).

\bibitem{Giacometti09b} A. Giacometti, F. Lado, J. Largo, G. Pastore, and F. Sciortino, J. Chem. Phys. \textbf{131}, 174114 (2009).

\bibitem{Giacometti10} A. Giacometti, F. Lado, J. Largo, G. Pastore, and F. Sciortino, J. Chem. Phys. \textbf{132}, 174110 (2010)

\bibitem{Lado82} F. Lado, Phys. Lett. \textbf{89}A, 196 (1982).

\bibitem{Lado82a} F. Lado, Mol. Phys. \textbf{47}, 283 (1982).

\bibitem{Lado82b} F. Lado, Mol. Phys. \textbf{47}, 299 (1982).

\bibitem{Lado95} F. Lado, E. Lomba, and M. Lombardero, J. Chem. Phys. \textbf{103}, 481 (1995).

\bibitem{Zwanzig54} R. Zwanzig, J. Chem. Phys. \textbf{22}, 1420 (1954)

\bibitem{Barker67} J.A. Barker and D. Henderson, J. Chem. Phys. \textbf{47}, 2856 (1967)

\bibitem{Gogelein08} C. G\"ogelein, G. N\"agele, R. Tuinier, T. Gibaud, A. Stradner, and P. Schurtenberger, J. Chem. Phys. \textbf{129}, 085102 (200

\bibitem{Gogelein12} C. G\"ogelein, F. Romano, F. Sciortino, and A. Giacometti, J. Chem. Phys. \textbf{136},094512 (2012)

\bibitem{Doi86} M. Doi and S.F. Edwards, \textit{Theory of Polymer Dynamics} (Oxford Univ. Press 1986)

\bibitem{Lowen94} H. L\"owen, Phys. Rep. \textbf{237}, 249 (1994)

\bibitem{Hynninen07} A.P. Henninen, J.H.J. Thijssen, E.C.M. Vermolen, M. Dijskra, and A. Van Blaaderen, Nat. Mater.
\textbf{3}, 593 (2007)

\bibitem{Hansen86} J. P. Hansen and I. R. McDonald, \textit{Theory of Simple Liquids} (Academic, New Yor

\bibitem{Labik85} S. Lab\'{i}k, A. Malijevsk\'{y}, and P. Vo\v{n}ka, Mol. Phys. \textbf{56}, 709 (1985).

\bibitem{Gray84} C. G. Gray and K. E. Gubbins, \textit{Theory of Molecular Fluids, Vol. 1: Fundamentals} (Clarendon, Oxford, 1984).

\bibitem{Vega92} L. Vega, E. de Miguel, L. F. Rull, G. Jackson, and I. A. McLure, J. Chem. Phys. \textbf{96}, 2296 (1992)

\bibitem{Liu05} H. Liu, S. Garde, and S. Kumar, J. Chem. Phys. \textbf{123}, 174505 (2005)

\bibitem{Sciortino09} F. Sciortino, A. Giacometti, and G. Pastore, Phys. Rev. Lett. \textbf{103} , 237801 (2009).

\bibitem{Sciortino10} F. Sciortino, A. Giacometti, and G. Pastore, Phys. Chem. Chem. Phys. \textbf{12}, 11869 (2010).

\bibitem{Henderson71} D. Henderson and J.A. Parker, in \textit{Physical Chemistry, an advanced treatise} Vol VIIIA page 377 (1971)

\bibitem{Barker76} J.A. Barker and D. Henderson, Rev. Mod. Phys. \textbf{48}, 587 (1976)

\bibitem{Henderson80} D. Henderson, O. H. Scalise, and W. S. Smith, J. Chem. Phys. \textbf{72}, 2431 (1980) 

\bibitem{Likos94} C.N. Likos, Zs T. N\`emeth, and H. L\"{o}wen, J. Phys: Condens. Matter \textbf{6}, 10965 (1994)

\bibitem{Weeks71} J.D. Weeks, D. Chandler, H.C. Andersen, J. Chem. Phys. \textbf{54}, 5237 (1971)

\bibitem{Andersen76} H. C. Andersen, D. Chandler and J.D. Weeks, Adv. Chem. Phys. \textbf{34}, 105 (1976)

\bibitem{Chandler83} D. Chandler, J.D. Weeks, and H.C. Andersen, Science \textbf{220}, 787 (1983)

\bibitem{Chandler72} D. Chandler and H.C. Anderson, J. Chem. Phys. \textbf{57}, 1930 (1972)

\bibitem{Bianchi11} E. Bianchi, R. Blaak and C. N. Likos, Phys. Chem. Chem. Phys. \textbf{13}, 6397 (2011)

\bibitem{Fantoni11} R. Fantoni, A. Giacometti, F. Sciortino, and G. Pastore, Soft Matter \textbf{7}, 2419 (2011)

\bibitem{Reinhardt11} A. Reinhardt, A. J. Williamson, J. P. K. Doyle, J. Carrete, L. M. Varele, and A. A. Louis,
J. Chem. Phys. \textbf{134}, 104905 (2011)

\bibitem{Tavares10} J.M. Tavares, P. I. C. Teixeira, M. M. Telo de Gama, and F. Sciortino, J. Chem. Phys. \textbf{132},
234502 (2010).

\bibitem{Bianchi06} E. Bianchi, J. Largo, P. Tartaglia, E. Zaccarelli, and F. Sciortino, Phys. Rev. Lett. \textbf{97}, 168301 (2006)


\end{thebibliography}

\end{document}